\def\ltsima{$\; \buildrel < \over \sim \;$}
\def\lsim{\lower.5ex\hbox{\ltsima}}
\def\gtsima{$\; \buildrel > \over \sim \;$}
\def\gsim{\lower.5ex\hbox{\gtsima}}
\documentstyle[12pt,aasms4]{article}
\input psfig.sty

\begin{document}

\title{SIGMA and XTE observations of the soft X-ray transient XTEJ1755-324
}
\author{P. Goldoni\altaffilmark{1,2}, M. Vargas\altaffilmark{2},
         A. Goldwurm\altaffilmark{1}, J. Paul\altaffilmark{1},
         V. Borrel\altaffilmark{3}, E. Jourdain\altaffilmark{3},
         L. Bouchet\altaffilmark{3}, J.-P. Roques\altaffilmark{3},
         M. Revnivtsev\altaffilmark{4}, E.Churazov\altaffilmark{4,5},
         M. Gilfanov\altaffilmark{4,5}, R. Sunyaev\altaffilmark{4,5},
         A.Dyachkov\altaffilmark{4}, N. Khavenson\altaffilmark{4},
         I.Tserenin\altaffilmark{4}, N. Kuleshova\altaffilmark{4}}
\affil{$^1$ CEA/DSM/DAPNIA/SAp, CEA-Saclay, F-91191 Gif-sur-Yvette, France}
\affil{$^2$
INTEGRAL Science Data Center, 16 Chemin d'Ecogia, CH-1290 VERSOIX, Switzerland}
\affil{$^3$ 
Centre d'Etude Spatiale des Rayonnements, 9 Avenue du Colonel Roche, BP 4346, 31029 Toulouse Cedex France}
\affil{$^4$ 
Space Research Institute, Profsouznaya 84/32, Moscow 117810, Russia}
\affil{$^5$ 
Max-Planck-Institut f\"ur Astrophysik, Karl-Schwarzschild-Str. 1,85740 Garching
bei Munchen, Germany}

\authoremail{paolo@discovery.saclay.cea.fr}

\begin{abstract}

\noindent We present observations of the X-ray transient XTEJ1755-324
performed during summer 1997 with the XTE satellite and with the SIGMA
hard X-ray telescope onboard the GRANAT observatory. The source was
first detected in soft X-rays with XTE on July 25 1997 with a rather
soft X-ray spectrum and its outburst was monitored in soft X-rays up
to November 1997. On September 16 it was first detected in hard X-rays by
the French soft $\gamma$ ray telescope SIGMA during a
Galactic Center observation. The flux was stronger on September 16
and 17 reaching a level of $\sim$ 110 mCrab in the 40-80 keV energy band.
On the same days the photon
index of the spectrum was determined to be $\alpha$ =$-$2.3$\pm$ 0.9
(1$\sigma$ error) while the 40-150 keV luminosity was $\sim$ 8 $\times$ 10$^{36}$ erg s$^{-1}$ for a distance of 8.5 kpc. SIGMA and XTE results on 
this source indicate that this source had
an ultrasoft-like state during its main outburst and a harder
secondary outburst in September. These characteristics make the source
similar to X-Nova Muscae 1991, a well known black hole candidate.

\end{abstract}


\section{Introduction}

\noindent The X-ray transient XTEJ1755-324 was discovered by the RXTE/ASM
on 25 July 1997 (Remillard et al. 1997) in the Galactic center region.
On the 29th of July it was observed with
the pointed instruments PCA (sensitive in the 2-30 keV energy band) and
HEXTE (20-100 keV band) and its precise position was determined to be 17h52m12sec, -32$^{\circ}$ 28' 12" (B 1950, uncertainty 1 arcminute) while
its flux was 170 mCrab in the 2-12 keV band (1 mCrab in the 2-12 keV band corresponds to 4.3 $\times$ 10$^{-11}$ erg cm$^{-2}$ s$^{-1}$).
The observed PCA spectrum was fitted with a multicolor disk blackbody with 
T$\sim$0.7 keV, and a hard tail extending up to 20 keV (Remillard et al. 1997).

\noindent The outburst of XTEJ1755-324 was followed by the RXTE/ASM in
the 2-12 keV energy band in the following months (see
http://space.mit.edu/XTE/). After the initial fast rise, the source flux
decayed in an exponential way and the source was detected up to the end of 
November. During all the observed outburst no type I X-ray burst
or pulsation, the most certain observational signatures of neutron stars
were reported from this source (P. Ubertini, private communication).
This light curve resembles the one of well
known X-ray Novae like X-Nova Muscae 1991 (Ebisawa et al., 1994) which is
also a good black hole candidate on the basis of its mass function (Mc Clintock et al., 1992, Orosz et al., 1996).
 
\noindent No optical counterpart has been proposed for XTEJ1755-324 while
a radio search performed on the 17-18 August with the Australia Telescope
Compact Array (Ogley et al., 1997) gave no detection up to a limit of
0.2 mJy (1384 MHz) and 0.3 mJy (2496 MHz).

\noindent We have carried out archival searches in the XTE 1 arcminute error box in the
ROSAT All Sky Survey Bright Source Catalog and we obtained no detection 
resulting in an upper limit of 0.05 ROSAT/PSPC cts/s for the source quiescent
emission. This translates in a 0.1-2.4 keV upper limit of $\sim$ 7 $\times$
10$^{-12}$ erg cm$^{-2}$ s$^{-1}$ assuming a Crab-like spectrum and 
10$^{22}$ cm$^{-2}$ absorbtion column density.

\section{SIGMA Observations and results}

\noindent The French coded mask telescope SIGMA onboard the Russian GRANAT
observatory provides high resolution images in the hard-X/soft $\gamma$-ray
band from 35 keV to 1300 keV, with a typical angular resolution of 15'
and a 20 hour exposure sensitivity (1$\sigma$) of $\sim$ 20 mCrab in
the 40-150 keV band (Paul et al., 1991). (1 mCrab is corresponds to 8.0 $\times$
10$^{-12}$ erg cm$^{-2}$ s$^{-1}$ in the 40-80 keV energy band and to 6.9 
$\times$ 10$^{-12}$ erg cm$^{-2}$ s$^{-1}$ in the 80-150 keV energy band).
The position determination accuracy of
the instrument can be 3-5 arcmin for a 6 $\sigma$ source and $<$1 arcmin for a 
30 $\sigma$ one. The Fully Coded field of view (FCFOV)
of the instrument is a 4$^{\circ}$.7 x 4$^{\circ}$.3 rectangle while the
Extended Field of View (EXFOV) is a bigger rectangle 18$^{\circ}$.1 x 16$^{\circ}$.8.

\noindent The Fall 1997 Galactic Center campaign began the 16th of
September 1997 (MJD 50707), 52 days after the discovery of XTEJ1755-324
(MJD 50655). A hard X-ray source was localized by SIGMA during the first 
observing session with 6' accuracy (90 $\%$ error circle) at R.A.=17hh 52mm
21sec, Dec=$-$32$^{\circ}$ 27'02" (B1950) about 2' from RXTE position stated
above and slightly outside the fully coded FOV. The SIGMA telescope continued 
its observations of the Galactic Center for a total of 5 observations and
about 123 hours of effective observing time.

\noindent During the campaign the hard X-ray flux of XTEJ1755-324 declined (see Table 1 and Figure 1) passing from $\sim$ 110 mCrab to 39 mCrab in the 40-80 keV energy band.
We thus analysed in more detail the results of the first observation
which took place between the 16th and the 18th of September when
the source was stronger. As can be seen in Table 1 and in Figure 1,
the source flux was stronger during the first half of this observation (September 16-17).

\noindent To better illustrate this fact we show in Fig.2 two 40-80 keV
images of a 6.5$^{\circ}$ x 6.5$^{\circ}$ region of the Galactic Center
containing XTE J1755-324 and 1E1740.7-2942 which was less variable during
all the campaign. The image on the left is a sum of the first half of
the first observation, the second image is a sum of the second half.
It appears that XTEJ1755-324 was the
strongest source in the field during the first part of the observation
appearing as a $\sim$ 6 $\sigma$ source with a mean flux of about 110 mCrab
in the 40-80 keV energy band and of 80 mCrab in the 40-150 keV energy band
(see Table 1). The hard X-ray outburst however ended quickly and in the second half of this observation the source was below the 4$\sigma$ level (Fig 2, right).

\noindent In order to have the smallest errors, we thus extracted a spectrum of
XTEJ1755-324 from the first half of the first observation. The result is shown
in Figure 5, the source is clearly detected up to $\sim $ 150 keV. We fitted
the resulting spectrum with a power law, our best fit being obtained with a
photon index $\alpha$=$-$2.3$\pm$0.9 and a 40-150 keV energy flux of
1($\pm$ 0.2) $\times$ 10$^{-9}$ erg cm$^{-2}$ s$^{-1}$.

\noindent The source luminosity was L$_X$(40-150) keV
$\sim$ 8$\times$10$^{36}$ erg s$^{-1}$ assuming a distance of 8.5 kpc.
To make a comparison X-Nova Muscae 1991 and X-Nova Velorum 1993 had
40-150 keV luminosities L$_X \sim$ 10$^{35}$ - 10$^{36}$ erg s$^{-1}$
resectively 100 and 120 days after their main outbursts (Goldwurm et al., 1993,
Gilfanov et al., 1991, Goldoni et al., 1998).

\noindent To investigate the possibility of previous activity from the source
we reanalyzed the SIGMA database of Galactic Center observations.
The SIGMA telescope monitored this sky region from 1990 to 1997 making
regular observations lasting about 0.5-1 month two times a year. We summed
the images of all these obsevations and searched for excess flux at the
source position. We did not detect any significant signal and we obtained a 2$\sigma$
upper limit of $\sim$ 3 mCrab in the 40-150 keV energy band.

\begin{table}

\caption{SIGMA observation log of the whole Galactic Center observations in
September 1997 with fluxes of XTEJ1755-324.  Errors quoted are at 68 \% confidence level in one parameter. Upper limit is at 68 \% confidence level. The fluxes of each observing session are presented, while the first observing session has been divided in two.
in two parts and its sum.}

\label{Table1}
\[
    \begin{array}{cccccc}
    \hline
\noalign{\smallskip}
  ${\rm Session}$ & ${\rm Date} $ & ${\rm Exposure}$
  & 40-80 {\rm keV } & 80-150 {\rm keV }
  & 40-150 {\rm keV } \\
                  &       {\rm  }       &   {\rm (hours)}   & 
  { \rm Flux (mCrab)} & { \rm Flux (mCrab)} & { \rm Flux (mCrab)} \\

\noalign{\smallskip}
\hline
\noalign{\smallskip}
1 $-{\rm I}$ & 1997~${\rm September}$ ~16.4-17.3 & 14.7 & 114 \pm 26 & 61 \pm 30 & 83 \pm 20\\
\hline
\noalign{\smallskip}
1 $- {\rm II}$ & 1997~${\rm September}$ ~17.3-18.2 & 14.7 & 64 \pm 26 &  < 30 & 19 \pm 20 \\
\hline
\noalign{\smallskip}
1 $-{\rm I + II}$ & 1997~${\rm September}$ ~16.4-18.2 & 29.2 & 84 \pm 19 & 22 \pm 21 & 48 \pm 15\\
\hline
\noalign{\smallskip}
2 & 1997~${\rm September}$ ~18.7-20.2 & 24.4 & 68 \pm 21 & 39 \pm 22 & 51 \pm 16 \\
\hline
\noalign{\smallskip}
3 & 1997~${\rm September}$ ~20.3-22.2 & 14.7 & 28 \pm 27 & < 30 & 2 \pm 21 \\
\hline
\noalign{\smallskip}
4 & 1997~${\rm September}$ ~22.8-24.2 & 22.7 & 28 \pm 22 & 41 \pm 24 & 36 \pm 16 \\
\hline
\noalign{\smallskip}
5 & 1997~${\rm September}$ ~24.3-26.3 & 31.7 & 39 \pm 18 & 5.5 \pm 20 & 19 \pm 14 \\
\hline
\noalign{\smallskip}
{ \rm Sum} & 1997~${\rm September}$ ~16.4-26.3 & 122.7 & 53 \pm 9 & 21 \pm 10 & 34 \pm 7 \\
\noalign{\smallskip}
  \end{array} 
   \]
\end{table}

\begin{figure}

\centerline{\psfig{figure=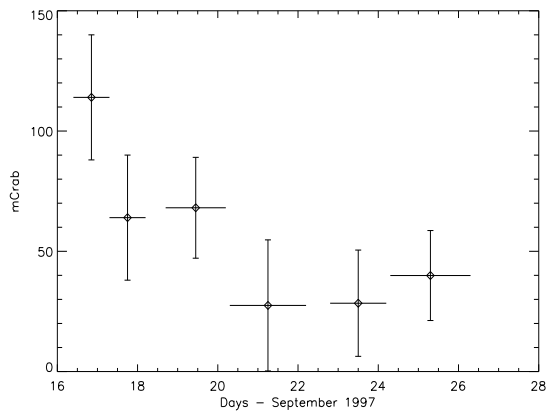,height=60mm,width=80mm}}

\caption[]{SIGMA 40-80 keV Light curves of XTEJ1755-324 in September 1997:
The data points are the average fluxes from each observing session with the
exception of the first two which are the fluxes of the two halves of the first
observing session. It is clearly visible a general flux decrease.
}

\label{Figure1}
\end{figure}


\begin{figure}[c]

\begin{picture}(300,200)(0,0)

 \put(50,200){\makebox(45,42)[tl]{%
\psfig{figure=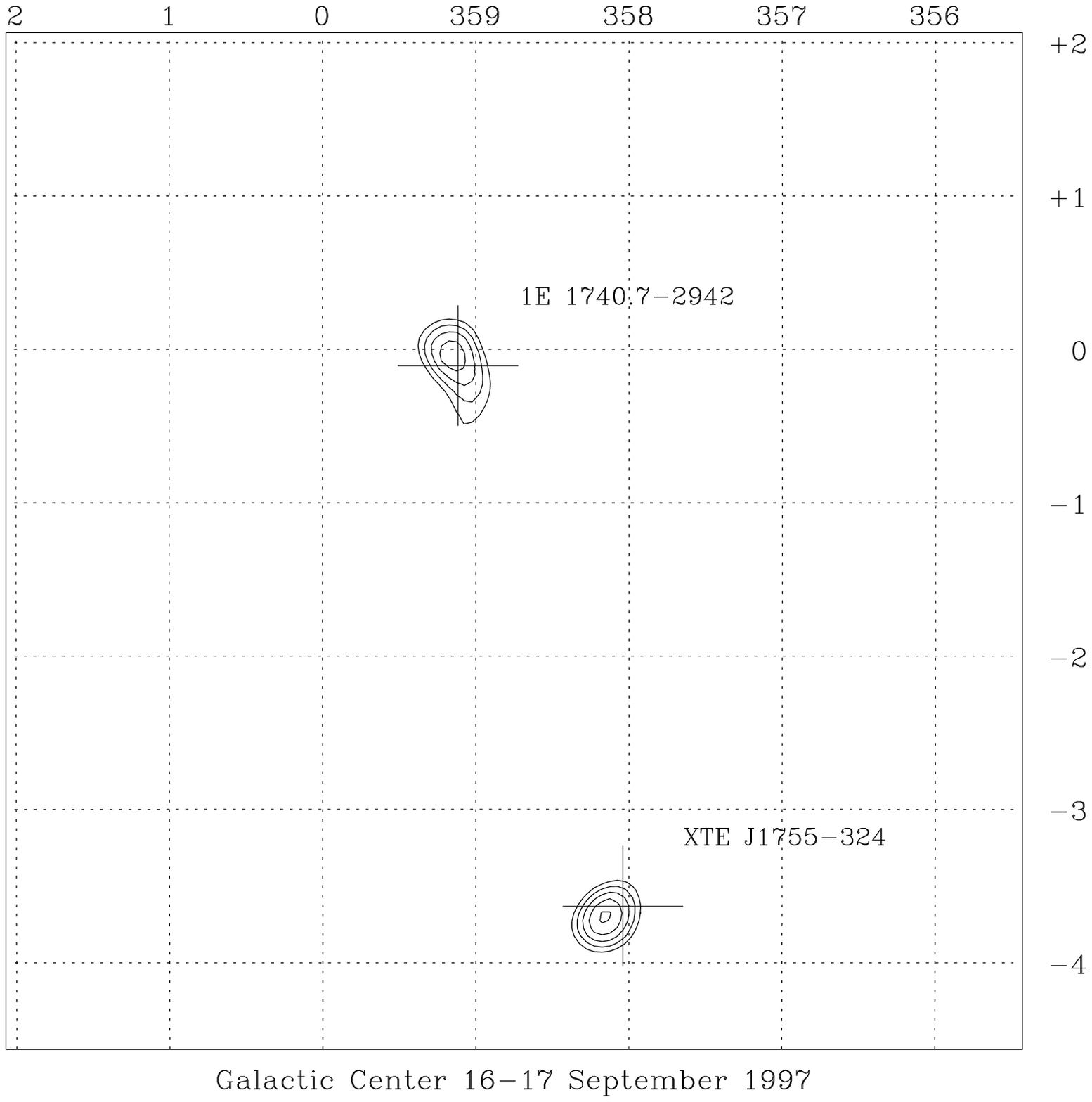,height=60mm,angle=90}}}
\put(250,200) {\makebox(45,42)[tl]{%
\psfig{figure=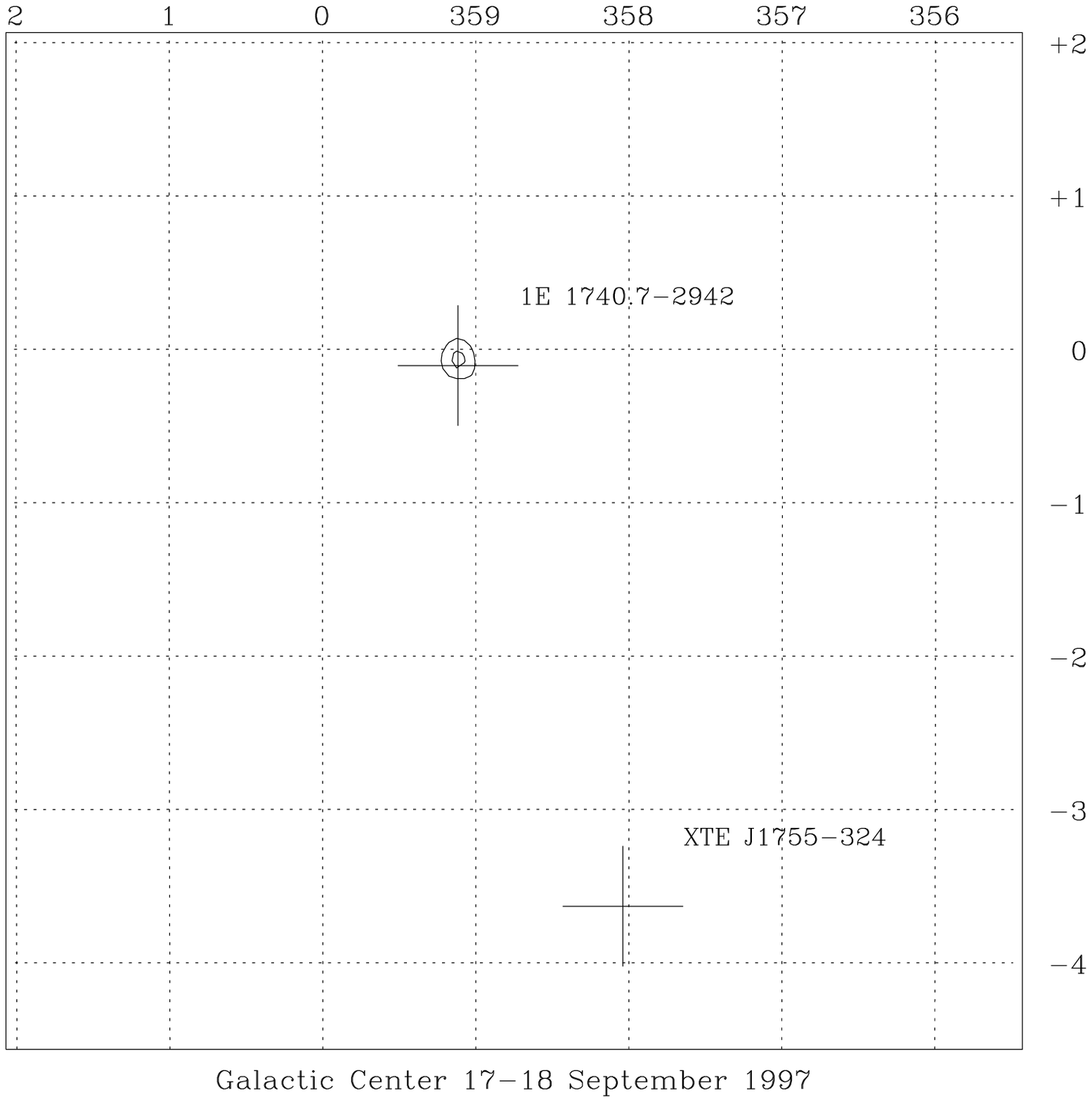,height=60mm,angle=90}}}

\put(0,0){\begin{minipage}[c]{\linewidth}

\caption[]{Contour image in the 40-150 keV energy band from the first three
exposures of the SIGMA observation of September 16-17 1997 (left) and for the
last three exposures (right). A 6.5 x 6.5 degrees region containing
XTE J1755-324 and 1E1740.7-2942 is shown. Confidence levels start from 
4.0$\sigma$ with 0.5$\sigma$ step. XTE J1755-324 stands out clearly as
the strongest source in the first image while it
disappears in the second one.}

\label{Figure2}
\end{minipage}}
\end{picture}
\end{figure}

\begin{figure}

\centerline{\psfig{figure=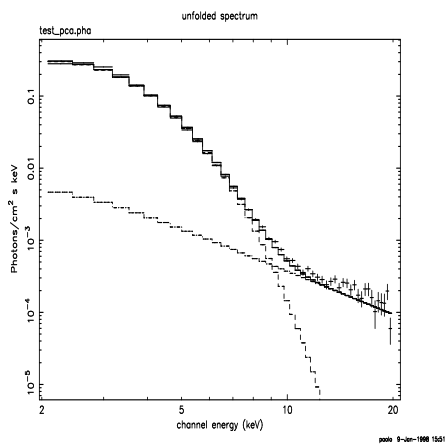,height=60mm,width=60mm,angle=-90}}
\caption{PCA 2-20 keV spectrum of XTEJ1755-324 on July 29th 1997, data points
are shown together with fitted spectrum. The power law component dominating at E$>$ 10 keV is clearly visible.}
\label{Figure3}
\end{figure}

\begin{figure}

\centerline{\psfig{figure=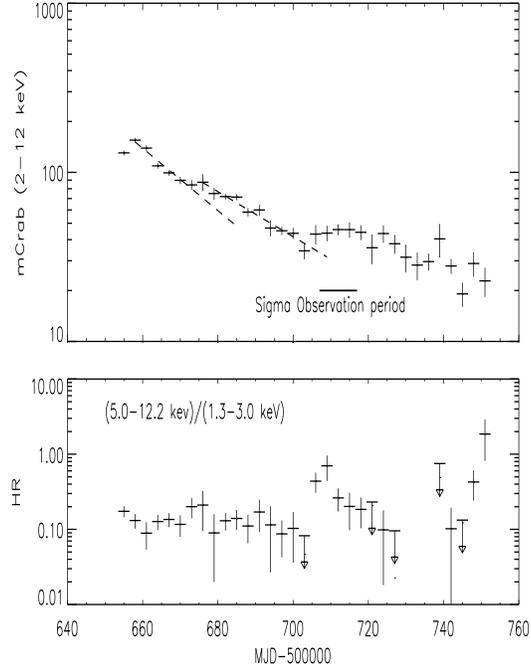,height=100mm,width=75mm}}
\caption{Upper panel:(1.3-12.2 keV) RXTE All Sky Monitor light curve of XTEJ1755-324. The horizontal line marks the epoch of SIGMA observations.
The dashed lines show linear fits before and after probable secondary maximum.
Lower panel: RXTE/ASM Hardness ratio (5.0-12.2 keV)/(1.3-3.0 keV) light
curve. The hardness ratio increases at the beginning of SIGMA observations
reaching a value of about 0.6 which roughly corresponds to a photon index
2.5. Some points are missing in the lower panel due to insufficient
statistics.}
\label{Figure4}
\end{figure}

\begin{figure}

\centerline{\psfig{figure=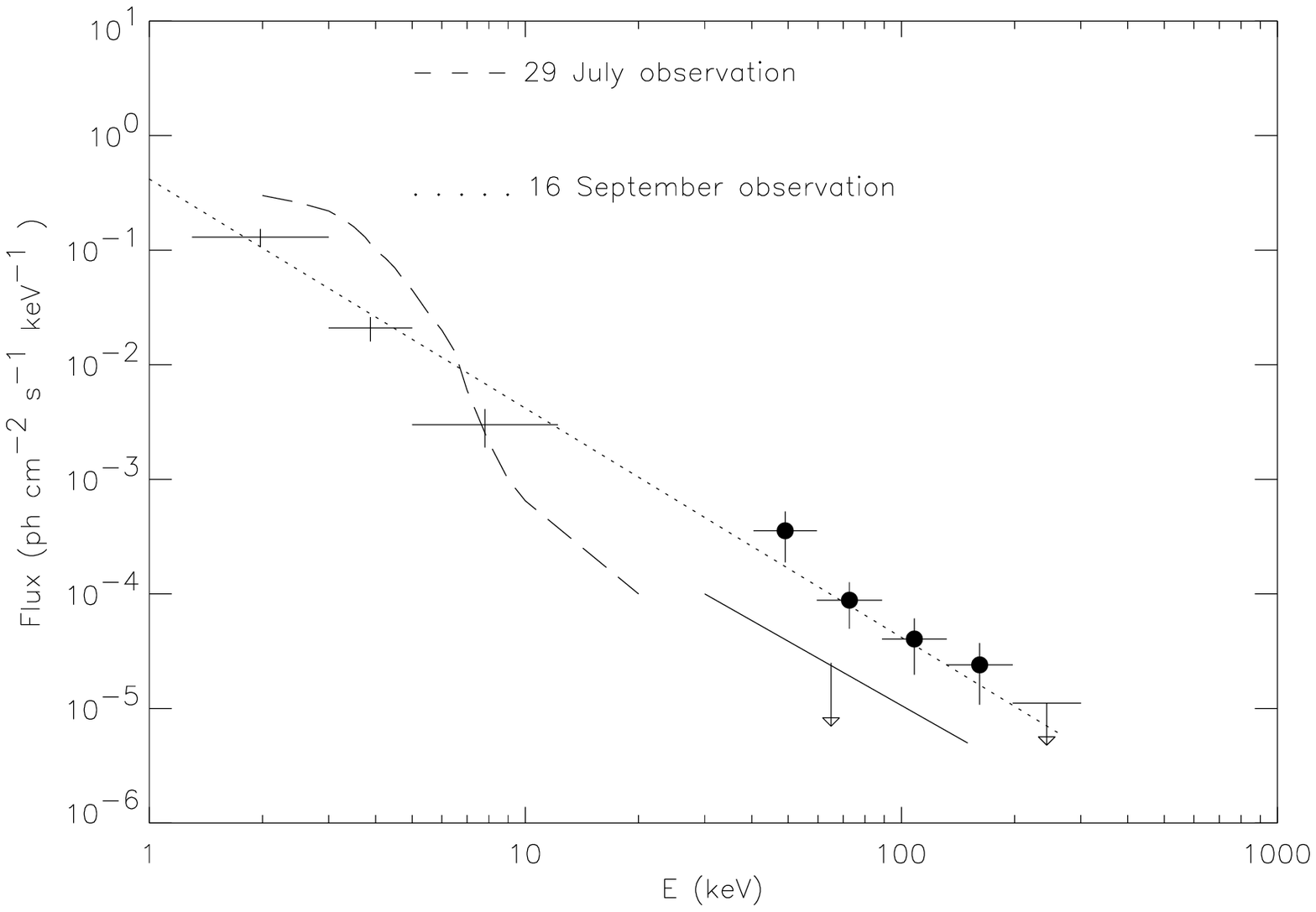,height=85mm,width=85mm}}
\caption{The 1-150 keV SIGMA/XTE spectrum of XTE J1755-324 taken
on September 16 1997 is shown together with a plot of the 29 July 1997 
PCA/HEXTE spectrum.  Filled points are SIGMA data, crosses are ASM data.
The dotted line is the best fit power law spectrum obtained from SIGMA data,
it is visible that this spectrum is also compatible with RXTE/ASM data.
The dashed line represents the source spectrum detected on July 29 1997, the
long upper limit stands for the HEXTE non detection. It is apparent that
the spectral shape was different at these two dates (see text for details).}

\label{Figure5}
\end{figure}

\section{RXTE Observations and results}

\noindent After its discovery, XTEJ1755-324 was observed by the
pointed instruments PCA and HEXTE on July the 29th 1997 and was
monitored by the All Sky Monitor during all of its outburst.

\noindent We reanalysed the public data of the pointed observations 
of the 29th of July and we found results compatible with those quoted by
Remillard et al. (1997) (see also Revnivtsev et al. 1998).
XTEJ1755-324 was clearly detected in the 2-20 keV
energy band and its emission could be fitted with a blackbody or a
multicolor disk blackbody (Makishima et al., 1986) with kT $\sim$ 0.7-0.8 keV 
and inner radius of the blackbody region R$_{in} \sim$ 36 $\times$
(cos $\theta$)$^{-1/2}$ km (assuming 8.5 kpc distance) plus a power law
extending up to about 20 keV with photon index $\alpha \sim -$2/$-$3
(Figure 3). We estimated a best fit hydrogen absorption column density 
of $\sim$1.5 $\times$10$^{22}$ cm$^{-2}$. The source was not detected with
HEXTE and we estimated a conservative 40-150 keV upper limit F$_x$
(40-150 keV) $\lsim$ 8$\times$ 10$^{-11}$ erg cm$^{-2}$s$^{-1}$ assuming
a power law spectrum with $\alpha \sim -$ 2. The 2-20 keV source luminosity
of the source was about 2 $\times$ 10$^{37}$ erg s$^{-1}$ at 
8.5 kpc distance. This spectrum characterised by an "ultrasoft" component
(kT $\lsim$ 1 keV) and a power law starting at about 10 keV was observed in
several X-ray Novae like X-Nova Muscae 1991 (Ebisawa et al., 1994) and
A0620-00 (Ricketts et al., 1975) at the beginning of their outbursts.
We also note that the upper limit on the hard X-ray flux is about 10-15 times
lower than the flux recorded by SIGMA on September 16-17 thus showing a
strong rise between the two observations.

\noindent  Further informations are provided by the XTE/ASM source light
curve in the 1.3-12.2 keV energy band. The light curve is shown in Figure
4 (upper panel), the count rate to energy conversion is based on the rule
that $\sim$ 75 ASM counts/sec are equal to 1 Crab in the 1.3-12.2 keV band.
In the lower panel of Figure 4 is plotted the ratio between the 5.0-12.2
keV and the 1.3-3.0 keV flux, this parameter roughly indicates the spectral 
shape in this band, hardness ratios 0.1 and 1.0 roughly correspond respectively 
to photon indices 4.0 and 2.0. In both graphics the plots
were produced using the FTOOLS task lcurve with a bin length of three days.
In the lower graph three points are not represented as the 5.0-12.2 keV
recorded flux was less than zero while in some others only upper limits are available.

\noindent Overall the source light curve has a typical FRED (Fast Rise Exponential Decay) shape with some interesting features. As it is shown
by superimposed log-linear fits, the exponential decay is probably
interrupted a first time around MJD 50680, i.e. $\sim $20 days after the outburst and then continues with a similar decay constant but with a
slightly higher intensity. This behaviour has been referred as a 'glitch'
and it has been seen in X-ray light curves of GROJ0422+32, A0620-00 and GRS1124-68 (Chen et al., 1997). The decay constant before and after the
proposed glitch are $\sim$ 24 $\pm$ 12 and $\sim $ 32 $\pm$ 20 days and are
thus compatible.

\noindent A bigger feature appears around MJD $\sim$ 50705 
where is clearly visible a stop in the flux decay and a possible flux rise
which lasts up to MJD $\sim$ 50725. After this event the source restarts its decline going under 20 mCrab at the end of November (MJD $\sim$ 50780). 
This type of event has been defined as a 'bump' and it has been seen in X-rays
in A0620-00 and in GRS1124-68 (Chen et al., 1997). We note that SIGMA observation were performed during the 'bump' on September 16-18 (MJD 50708-50710).
Moreover the ASM hardness ratio shows evidence of a spectral hardening
in the 1.3-12.2 keV band at the same time, (see Figure 4, lower panel).
In fact the average hardness ratio  of the source in the
period from MJD 10655 to MJD 10706 was $<$HR$>$=0.14$\pm$0.06 while in the days
from MJD 10706 to MJD 10712, it was $<$HR$>$=0.6$\pm$0.2.
The implied photon index evolves from $\sim$ 4 to $\sim$ 2.5,
indicating a rise in the flux at E $>$ 5 keV.

\noindent In Figure 5 we plotted the 1-150 keV ASM/SIGMA spectrum of
XTEJ1755-324 on September 16th along with the SIGMA spectral fit. It
can be seen that all this spectrum is compatible with a single power
law with photon index $\alpha \sim$ 2.3 and that no strong ultra soft
excess is  present. However we cannot rule out the presence of a weak
soft excess at low energies whose presence could be easily masked by the
strong hard component. In any case on that date the 2-12 keV X-ray flux was
about 40 mCrab, 4 times lower than at outburst peak while, as seen before,
40-150 keV x-ray flux was 10-15 times stronger than on 29th of July.
A possible interpretation for this behaviour is that a soft-to-hard state 
transition similar to the ones which happen in the well known black hole candidate Cyg X-1 took place
between the two observation. We must remark however that our data
allow also the existence of a weak ultrasoft component and that no firm conclusion can be taken on this point.

\section{Discussion}

\noindent XTEJ1755-324 is an x-ray transient which displayed spectral
evolution in detections at different epochs by RXTE
and GRANAT/SIGMA. We will show in the following that
the evidences point towards a classification of this source as
an X-ray Nova of the type of X-Nova Muscae 1991.

\noindent The outburst light curve recorded by RXTE/ASM is characterised
by a fast rise (2-3 days) and an exponential decay with time scale
of about 20-30 days with a possible secondary outburst $\sim$ 20 days after the
primary outburst and a strong 'bump' after about 45 days.
This behaviour has been observed in X-Nova Muscae 1991 and A0620-00
in the standard (1-10 keV) X-ray band (see for example Chen et
al., 1997). The spectrum at the outburst peak could be fitted with
a multicolor disk blackbody or a blackbody with kT $\lsim$ 1 keV and a
power law component which is again typical of these sources (Tanaka \& 
Shibazaki, 1996). In our hypothesis the source was then in a typical
"ultrasoft" state which is frequently recorded during first phases of
X-ray Novae outbursts.

\noindent The hard X-ray flux was rather faint at outburst peak
while it was $\gsim$ 10-15 times stronger on September 16th decaying in the
following days. In the meantime the 2-12 keV emission was undergoing a rather
long secondary outburst or bump whose evolution was apparently uncorrelated
with the hard X-ray one. The hard X-ray spectral index recorded on September
16-17 was $\alpha \sim -$ 2.3 $\pm$ 0.9 and the source was detected
above 100 keV. As shown in Figure 5, ASM and SIGMA data recorded on this date
are compatible with the hypothesis of a 1-150 keV spectrum described by a single
power law with $\alpha \sim$ 2.3. Moreover the ASM (5.0-12.2)/(1.3-3.0) keV
hardness ratio on that day indicates taht the soft X-ray spectrum was harder
than during the outburst maximum and its value indicates a spectrum with photon
index $\sim$ 2.5.
The errors in our spectra are too big to determine a spectral index typical
of a 'low' state (i.e. $\alpha \sim$ 2) or of a 'high' state ($\alpha \sim$ 3).
In any case the wide band spectral shape suggests that the source was the
source either had a weak ultrasoft component not detectable in our data or 
was in a spectral state analogous to the 'low' state of Black Hole candidates 
like Cyg X-1.
An analysis of short term variability behaviour would secure the identification
of the spectral state of the source, unfortunately, the source is too faint 
for timing analysis in SIGMA data while ASM data points do not have the
necessary timing resolution.

\noindent This behaviour characterised by a strong ultrasoft component
at the beginning of the outburst followed by a strong rise of the hard
X-ray part of the spectrum has been detected in the outburst of
X-Nova Muscae 1991,  (see e.g. Kitamoto et al. 1992, Ebisawa et al. 1994, 
Goldwurm et al. 1993). The X-ray Nova GRS 1009-45 also displayed this behaviour
(Sunyaev et al. 1994, Goldoni et al. 1998). This behaviour can be naturally
explained in bulk flow comptonization models (Titarchuk, Mastichiadis \& 
Kylafis, 1997, Laurent \& Titarchuk, 1998) as due to changes in mass accretion 
rate. In this scenario the high mass accretion rates happening during the 
primary outburst produce a strong soft X-ray flux which efficiently cools
down the Compton cloud around the disk. The Compton cloud becomes thus 
transparent to soft X-ray radiation and no thermal comptonization emission
is visible. The hard X-ray emission in this case is due to bulk motion comptonization happening closer to the black hole which produces a spectral
index $\sim -$ 2.8 (Laurent \& Titarchuk, 1998). In the context of these
models, the presence of the extended power law together with the ultrasoft
component is considered to be a black hole signature. In our interpretation
this is the spectral state that was observed by XTE on July 29th. Later in
the outburst, however, the mass accretion rate is lower and the optical
depth of the Compton cloud increases thus causing the appearence of the typical
thermal Comptonization spectrum which can be modeled with a power law with
spectral index $\sim -$ 2 in the energy range 1-150 keV compatible with the
spectrum SIGMA and XTE observed on September 16th. In our case however, a
weak ultrasoft component was possibly present together with the power law component and we cannot be sure of the origin of the emission on that date.

\noindent Unfortunately without an optical or radio counterpart we do
not have precise information on the source distance. The absorption
column density we estimated from RXTE/PCA observations is strongly
dependent on the spectrum. A possible distance estimate can be done in
our hypothesis that the source is an X-ray Nova, in that case its
outburst luminosity can be estimated to be near the Eddington luminosity
in the standard X-ray band (Chen et al., 1997). Taking into account the
absorption column density we quoted and
the relatively low flux at maximum, we tentatively suggest that the source
is in the Galactic Bulge. If this is the case, it would be the fourth
faint (F$_{max} <$ 1 Crab) Galactic Bulge hard X-ray transient detected by
SIGMA after GRS1737-31 (Trudolyubov et al. 1998, Cui et al., 1997),
GRS 1730-312 (Vargas et al., 1996, Trudolyubov et al., 1997), and GRS
1739-278 (Vargas et al., 1997).

\section{Conclusion}

\noindent We reported on the hard X-ray detection of XTEJ1755-324
with the SIGMA telescope. The general characteristics of its
light curve and its spectrum characterize it as an X-ray Nova.
The soft X-ray light curve with a Fast rise and an Exponential Decay is
typical of black hole candidates like X-ray Nova Muscae 1991 (Ebisawa
et al., 1994) and A0620-00 (Ricketts et al., 1975).
Its spectrum was dominated by a "ultrasoft"(kT $\lsim$ 1 keV)
component during the primary outburst and by a power law 
extending up to more than 100 keV during SIGMA September observations
50 days after the primary outburst. The general behaviour of the source during
the outburst, together with the absence of Type I X-ray bursts and pulsations, point us to suggest
that this source is a
black hole candidate which is likely in the Galactic Bulge like
other three faint transient sources detected by SIGMA.

\noindent This class of sources is a natural target of SIGMA thanks to its 
localization accuracy, unprecedented in this energy band. It has been proposed 
(Vargas et al. 1997) on the basis mainly of SIGMA data that these transients
reach a peak luminosity of $\sim$ 10$^{37}$ erg s$^{-1}$ in the hard X-ray
band. This would mean that, with its limited sensibility, SIGMA can detect
these transients only in a limited region of our galaxy, i.e. up to about 10 
kpc. To detect more transients of this type more sensitive instruments are
needed.

\noindent The IBIS (Imager on Board Integral Satellite) coded mask telescope
(Ubertini et al., 1997) has a predicted 40-150 keV 3$\sigma$ sensitivity of 
about 3.5 mCrab in a one day observation. IBIS will thus be able
to detect flaring transients at distances up to 30 kpc (peak flux about
12 mCrab for a 10$^{37}$ erg s$^{-1}$ luminosity, i.e. in
all the Galaxy. Thanks to IBIS source localization capability and to the low absorption in this energy band, it will be possible
to make a reliable, experimental, census of all the hard X-ray Novae in the Galaxy.

\acknowledgments  We acknowledge the paramount contribution of the SIGMA
Project Group of the CNES Toulouse Space Center to the overall success
of the mission. We thank the staffs of the Lavotchine Space Company, of
the Babakin Space Center, of the Baikonour Space Center, and the Evpatoria 
Ground Station for their unfailing support. This research has made use of data
obtained through the High Energy Astrophysics Research Center Online Service, 
provided by the NASA/Goddard Space Flight Center.

\end{document}